\documentclass[11pt,twoside]{article}
\usepackage{newpasp}

\usepackage{epsf}

\index{Smith, B. J.}
\index{Donahue, M.}
\index{Stocke, J.}

\begin{document}

\title{Interstellar Gas and Star Formation Near the Peculiar Radio Galaxy
NGC 4410A}
\author{Beverly J. Smith}
\affil{East Tennessee State University, Physics and Astronomy 
Department,
Johnson City TN  37604}
\author{Megan Donahue}
\affil{Space Telescope Science Institute, Baltimore
MD  21218}
\author{John Stocke}
\affil{University of Colorado, APS Dept, Box 391, Boulder CO  80309}

% A concise abstract is recommended.  Enter the text of the abstract in
% between the \begin{abstract} and \end{abstract} commands.  Do NOT
% include the word ``Abstract'' in your text; it is insterted
% automatically. Do NOT  make a paragraph break between \begin{abstract} 
% and the first line of the text of the abstract!  Abstracts are required 
% for all papers.

\begin{abstract}

We present new multi-wavelength
data for the NGC 4410 galaxy group.
This group contains the peculiar radio galaxy NGC 4410A,
which has a distorted double-lobed radio structure.
NGC 4410A is strongly interacting with three nearby galaxies,
and has a peculiar ring-like optical morphology, a number
of luminous H~II regions, as well
as tidal tails and bridges variously visible in optical,
HI, and X-ray light.  
The gravitational interaction
has affected the interstellar matter, the star formation rate,
and the radio lobes in this system.

\end{abstract}

% Include keywords if you wish. The keywords.apj file, found on aas.org 
% in the pubs/aastex-misc directory, contains a list of keywords used 
% with the ApJ and Letters.  

%%\keywords{infrared: galaxies -- galaxies: nuclei -- galaxies: starburst}

% That's it for the front matter.  On to the main body of the paper.

\section{Introduction}

The large radio lobes of radio
galaxies both affect and are affected by their
environment.  They may be distorted by motion
through an intracluster medium
or by an encounter with interstellar matter in the
host galaxy or a companion
galaxy.
In addition, impacts between jets and interstellar matter
may trigger star formation.
To investigate the interplay between radio lobes and the
interstellar matter in radio galaxies, detailed studies
of the interstellar matter in nearby radio galaxies
are essential.  In this paper, we present a study
of the interstellar matter in the nearby low
luminosity radio galaxy NGC 4410A.

\section{A Multi-Wavelength View of the NGC 4410 Group}

NGC 4410A is part of a small group containing a dozen
known members.  
NGC 4410A has a prominent bulge surrounded by a ring-like
or loop-like structure (Figure 1a).  East and southeast
of the nucleus, luminous H~II regions
are detected (Figure 1b).
There is an HI tail extending
to the southeast of the NGC 4410A+B pair, coincident with
a stellar tail (Figure 2).  
NGC 4410A was also detected in CO (1 $-$ 0),
yielding
M$_{H_2}$ $\sim$ 4 $\times$ 10$^9$ M$_{\sun}$.
NGC 4410A has two 
distorted radio lobes;
the southeastern lobe 
is superposed on the optical/HI tidal tail.
X-ray maps
show a 2$'$ long `tail'
coincident
with the NGC 4410A+B/NGC 4410C stellar bridge.

\section{Conclusions}

The NGC 4410 group shows
clear evidence for gravitational interactions between
galaxies as well as evidence for an external force acting upon
the radio lobes.  
The peculiar ring-like structure of
NGC 4410A suggests that a near-head-on
collision with NGC 4410B has occurred.
We suggest that the distortion of the radio lobes was caused
by ram pressure from the interstellar medium,
due to the motion of this gas relative to the radio
source during a gravitational interaction 
between galaxies.
The X-ray extension may
be due to shocked tidal gas or
intragroup gas coincidently superposed
on the stellar bridge.
The 
star formation may have been induced by the collision or 
by a jet/interstellar medium impact.

{\bf Captions}

1a) a red continuum image of NGC 4410A (right)
and its companion NGC 4410B.  1b) an H$\alpha$+[N~II] image
of NGC 4410A+B.  The field of view is $\sim$1$'$.

2) A VLA D Array 21 cm HI map of the inner
part of the NGC 4410 group, superposed on an R-band image
from the SARA 0.9m telescope.  The field of view is 6$'$ $\times$
5$'$.

% For examples on including figures, see the file vla2000_sample.ps
% at http://www.nrao.edu/vla2000/proceedings/. 
% For examples of figures, equations or tables, please see the file
% vla2000_man.ps at the same site. Also available as
% newpaspman.ps at http://www.aspsky.org/pubs/authors.html

% comment this out if you want to include acknowledgements

%\acknowledgements

%\begin{references}
{\small

%\reference Smith, B. J. 2000, \apj, in press.

}
%\end{references}

\end{document}